\documentclass[aps,prl,twocolumn,showpacs,superscriptaddress,amsmath,amssymb,longbibliography]{revtex4-1}
\usepackage{graphicx}
\usepackage{color,soul}

\begin{document}

\title{Fundraising and vote distribution: a non-equilibrium statistical approach}

\author{H. P. M. Melo} \email{hygor.piaget@ifce.edu.br}
\affiliation{Instituto Federal de Educac\~ao, Ci\^encia e Tecnologia do
Cear\'a, Avenida Des. Armando de Sales Louzada, 62580-000 Acara\'u, Cear\'a,
Brazil}
  
\author{N. A. M. Ara\'ujo} \email{nmaraujo@fc.ul.pt}
\affiliation{Departamento de F\'{\i}sica, Faculdade de Ci\^{e}ncias,
Universidade de Lisboa, 1749-016 Lisboa, Portugal} \affiliation{Centro de
F\'{i}sica Te\'{o}rica e Computacional, Universidade de Lisboa, 1749-016
Lisboa, Portugal} \affiliation{Departamento de Fisica, Universidade Federal
do Cear{\'a}, 60451-970 Fortaleza, Cear{\'a}, Brazil}

\author{J. S. Andrade Jr.} \email{soares@fisica.ufc.br}
\affiliation{Departamento de Fisica, Universidade Federal do Cear{\'a},
60451-970 Fortaleza, Cear{\'a}, Brazil}


\begin{abstract}
The number of votes correlates strongly with the money spent in a campaign, but
	the relation between the two is not straightforward. Among other
	factors, the output of a ballot depends on the number of candidates,
	voters, and available resources. Here, we develop a conceptual
	framework based on Shannon entropy maximization and Superstatistics to
	establish a relation between the distributions of money spent by
	candidates and their votes.  By establishing such a relation, we
	provide a tool to predict the outcome of a ballot and to alert for
	possible misconduct either in the report of fundraising and spending of
	campaigns or on vote counting. As an example, we consider real data
	from a proportional election with $6323$ candidates, where a detailed
	data verification is virtually impossible, and show that the number of
	potential misconducting candidates to audit can be reduced to only
	nine. 
\end{abstract}

\maketitle

In an effort towards fair electoral processes, regulations and reforms are
constantly on the agenda of many countries around the world~\cite{UN2009}. To
avoid that the decision-making process is dominated by wealth and influence,
the most pertinent processes to legislate are arguably fundraising and
spending~\cite{IDEA2002}. Different countries have different rules, but in
general, candidates and parties are the ones that report on the financial
details of their own campaigns, what raises obvious doubts over the veracity of
the reported data. As the number of collected votes correlates with the money
spent in the campaign~\cite{Melo18}, establishing a quantitative relation
between the distribution of votes and financial resources among the candidates
is instrumental to raise flags about possible misconduct.
\begin{figure}
\includegraphics[width=0.85\columnwidth]{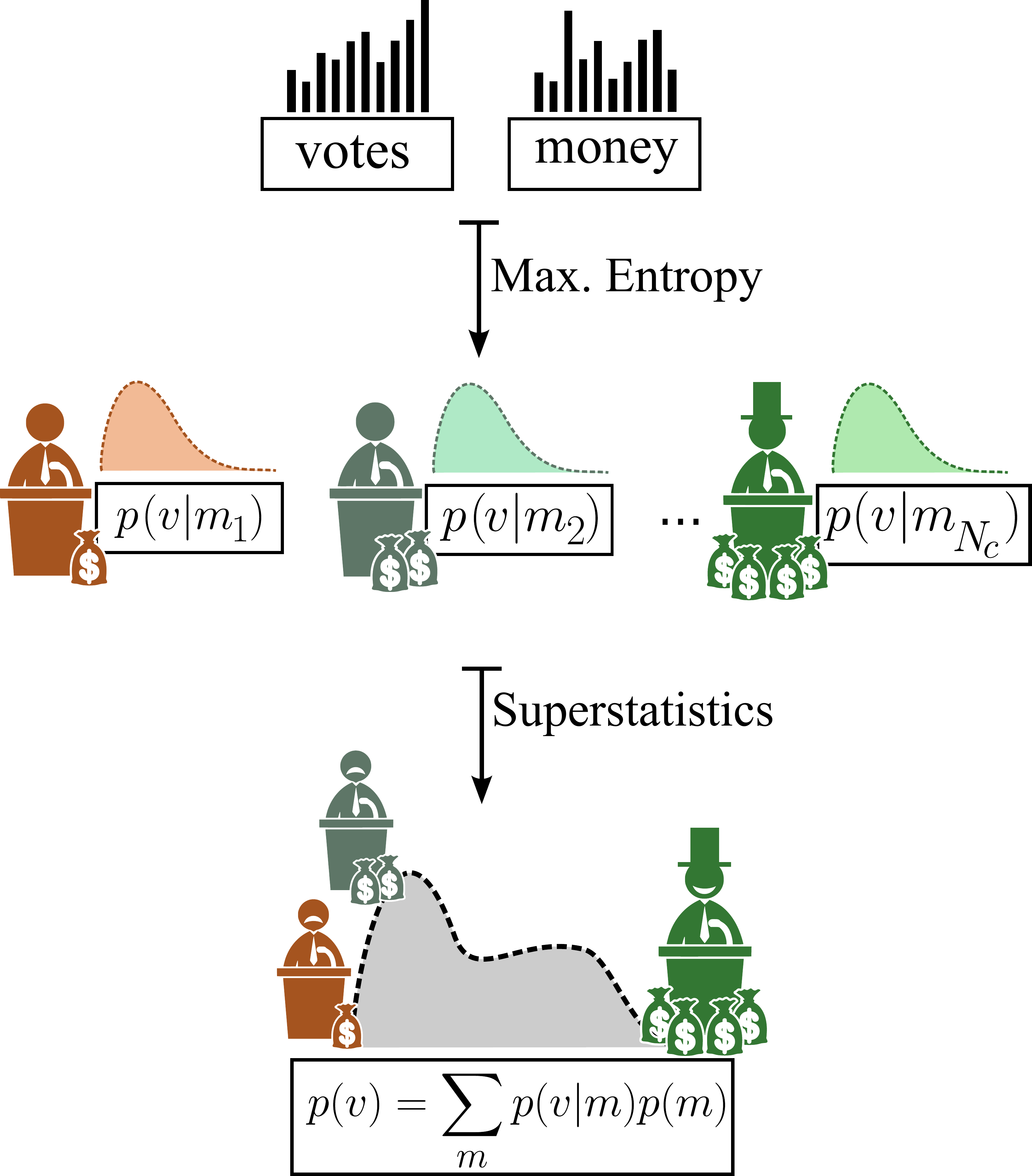}
\caption{By employing the principle of maximum entropy under the constraints of
	a fixed number of voters and candidates, we derive the conditional
	probability $p(v|m_i)$, that a candidate $i$ receives $v$ votes,
	provided that $i$ spends an amount of money $m_i$ in the campaign.
	Since the amount of money spent usually differ from candidate to
	candidate, the final distribution of votes should depend on the
	distribution of money spent. A formalism based on
	Superstatistics~\cite{Beck2003} is then used to establish a relation
	between these two distributions.~\label{fig::scheme}}
\end{figure}

Within some regulated boundaries, several individuals or institutions can
contribute financially to a campaign. The value of the contribution is very
subjective, depending on their interests and on the economic and political
conjecture~\cite{jacobson1978,Morton1992,gerber2004,Gordon2007}. Thus,
predicting the distribution of funds raised and money spent in a campaign from
``first principles'' is likely a hopeless endeavor, challenging the
verification of the reported data. In sharp contrast, the distribution of votes
among candidates is well studied. It is known to differ for proportional and
plural elections, and to depend on the country, number of candidates, and money
spent in
campaigns~\cite{Costa1999,Costa2003,Castellano2009,mantovani2011,mantovani2013,bokanyi2018}.
Different models were developed to explain this
distribution~\cite{Melo18,Moreira2006,Araujo2010,FernandezGracia2014,Calcao2015,Borghesi2012,Fortunato2007}
as well as methodologies to identify vote-counting
irregularities~\cite{lehoucq2003,alvarez2009,deckert2011,Klimek2012,beber2012,enikolopov2013}.
Here we propose an approach based on the Shannon entropy maximization and
Superstatistics to disclose a relation between the distribution of financial
resources declared by candidates and the distribution of their votes in
proportional elections.

\begin{figure*}[t]
\includegraphics[width=\textwidth]{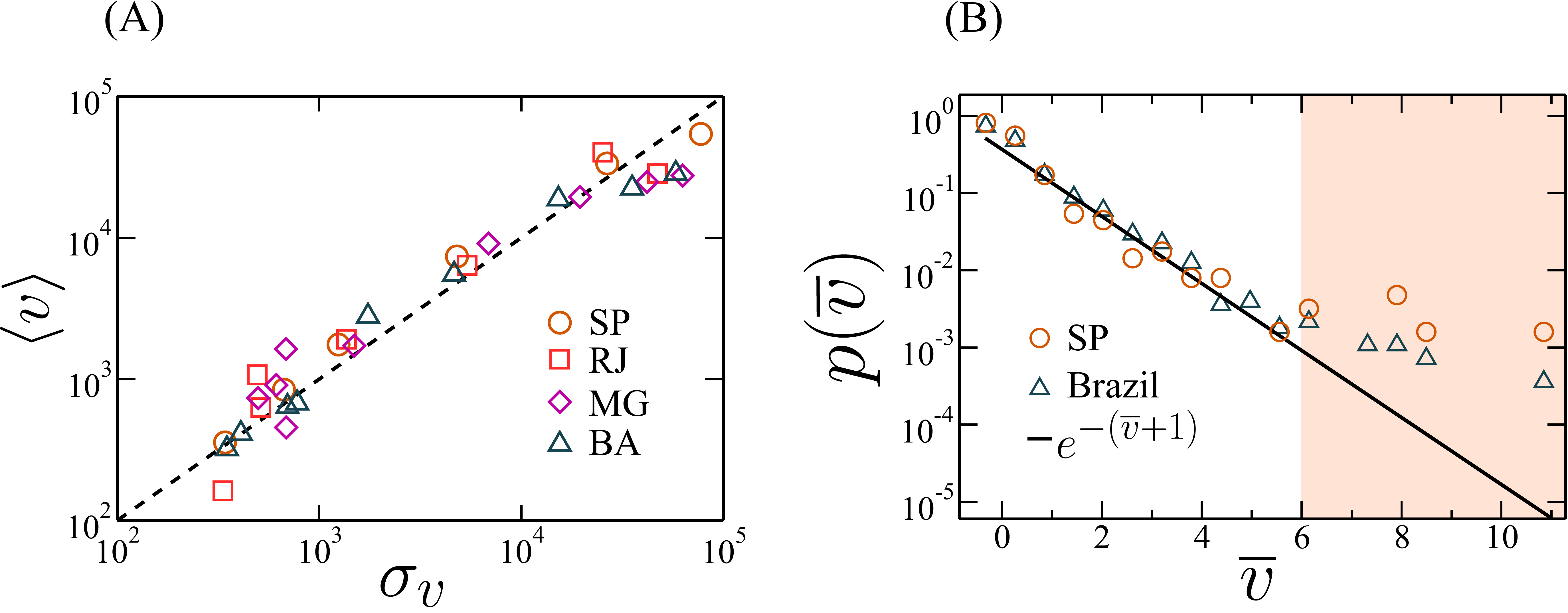}
\caption{Empirical data for the 2014 election for federal deputies in Brazil,
	which counted $6323$ candidates, roughly $140$ million voters, and more
	than $280$ million dollars of total investment in campaigns. (A) The
	average number as a function of the standard deviation of votes per
	candidate at the state level, for the top four populated Brazilian
	states, namely, S\~ao Paulo (SP, circles), Rio de Janeiro (RJ,
	squares), Minas Gerais (MG, diamonds), and Bahia (BA, triangles). For
	each state, candidates were grouped by the amount of money that they
	officially declared to have spent in their campaigns. The dashed line
	corresponds to a linear behavior. (B) Distribution of rescaled number
	of votes $\bar{v}=(v-\langle v \rangle)/\sigma_v$ for S\~ao Paulo (blue
	triangles) and the entire country of Brazil (black circles), where
	$\langle v \rangle$ and $\sigma_v$ are the average and standard
	deviation of the number of votes received by the candidates in the same
	interval (logarithmic binning) of money spent. The (red)-dashed line
	corresponds to $p(v)=\exp(-\bar{v}-1)$, as predicted by
	Eq.~\eqref{eq::distribution}, if we assume $Z(m)=\mu$. Following the
	prediction given by Eq.~\eqref{eq::distribution} from our theoretical
	approach, the distribution of votes for more than $99\%$ of the
	candidates follows an exponential distribution. However, it is
	remarkable that the number of candidates with votes that deviate more
	than $6\sigma_v$ (highlighted region) from the average is higher than
	expected, suggesting the existence of outliers.~\label{fig::data1}}
\end{figure*}

Given a certain amount of money $m_i$ spent by a candidate $i$ in the campaign,
the conditional probability for $i$ to receive $v$ votes is $p(v|m_i)$. Since
the money spent is heterogeneously distributed among candidates, the
probability $p(v)$ that a candidate receives $v$ votes is given by,
\begin{equation}\label{eq::superstatistics}
p(v)=\sum_{m=0}^{m_\mathrm{max}} p(v|m) p(m) \ \ , 
\end{equation}
where $p(m)$ is the probability that a candidate spends an amount of money $m$
in the campaign and $m_\mathrm{max}$ is the maximum amount of money that can be
spent (see Fig.~\ref{fig::scheme}). For simplicity, we have considered that $m$
is a discrete variable and a multiple of $\Delta m$, where $\Delta m$ is the
``price of a vote''. Equation~\eqref{eq::superstatistics} is the basis of
Superstatistics for non-equilibrium systems~\cite{Beck2003}, a theoretical
framework developed to describe the thermal fluctuations of an ensemble of
particles at different effective thermostat temperatures and consequently
different weights for each configuration.  Analogously, in an election, the
amount of money spent differs from candidate to candidate and thus also the
probability that they receive a certain number of votes. As a consequence, the
variable $m$ is the analogue for elections of the thermostat temperature in a
thermal system.
\begin{figure*}[t]
\includegraphics[width=\textwidth]{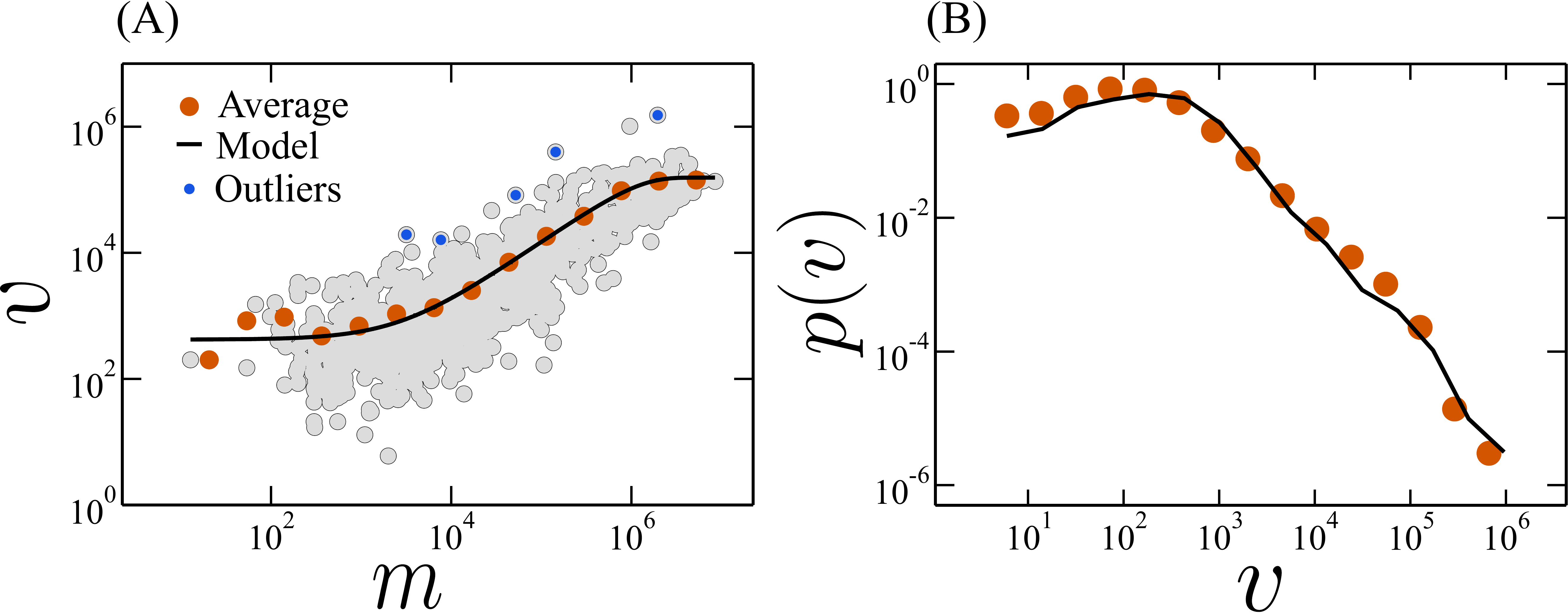}
\caption{Results for the 2014 election for federal deputies in the state of
	S\~ao Paulo, Brazil, with $1364$ candidates, more than $32$ million
	voters, and about $58$ million dollars of total investment in the
	campaigns. (A) Number of votes as a function of the money spent in the
	campaign for all candidates (gray dots) and the average value within
	each money group (blue-dark dots).  The (red-)solid line is obtained by
	a least-squares fit of Eq.~(\ref{eq::average_vote}) to the average
	data, taking $\Delta m$ as a free parameter. The black circles are
	outliers, which we defined as the candidates with a number of votes
	that deviate more than $6\sigma_v$ from the average. (B) Distribution
	of the number of votes per candidate. The (red) dots were obtained from
	the data and the solid line was calculated from the distribution of
	money spent in the campaign. Precisely, the solid line is obtained by
	randomly assigning a number of votes $v$ for each candidate from the
	distribution given by Eq.~\eqref{eq::distribution}, where $m$ is the
	amount of money officially declared to have been spent in the campaign.
	The obtained curve is remarkably consistent with the empirical data
	over more than five orders of magnitude.~\label{fig::data2}}
\end{figure*}

To calculate $p(v|m)$, let us consider a proportional election with $N_c$
candidates and $N_v$ voters. Based on the principle of maximum
entropy~\cite{jaynes1957}, $p(v|m)$ should maximize the Shannon entropy,
\begin{equation}\label{eq::Shannon}
S=-\sum_{i=1}^{N_c}\sum_{v=v_0}^{m_i/\Delta m} p(v|m_i) \ln [p(v|m)] \ \ ,
\end{equation}
where $v_0$ and $m_i/\Delta m$ are the minimum and maximum number of votes that
the candidate $i$ can receive. For simplicity, hereafter we assume that $v_0$
is the same for all candidates. At this point, two constraints need to be
imposed, as both the number of candidates $N_c$ and voters $N_v$ are fixed (see
Fig.~\ref{fig::scheme}). In this way, the first constraint is then,
\begin{equation}\label{eq::constraint1}
\sum_{i=0}^{N_c}\sum_{v=v_0}^{m_i/\Delta m} p(v|m_i) = N_c \ \ ,
\end{equation}
which ensures the normalization of $p(v|m)$, while the second one is,
\begin{equation}\label{eq::constraint2}
\sum_{i=0}^{N_c}\sum_{v=v_0}^{m_i/\Delta m} v p(v|m_i) = N_v \ \ .
\end{equation}
By maximizing $S$ subjected to
Eqs.~\eqref{eq::constraint1}~and~\eqref{eq::constraint2}, we obtain
\begin{equation}\label{eq::distribution}
p(v|m)=\frac{1}{Z(m)}e^{-\mu v} \ \ ,
\end{equation}
where $Z(m)$ is a normalization factor that depends on $m$ and it is the
analogue of the partition function in a thermal system, given by,
\begin{equation}\label{eq::normal}
Z(m)=\frac{e^{\mu(1-v_0)}-e^{-\mu m/\Delta m}}{e^\mu-1} \ \ ,
\end{equation}
where $\mu$ is the Lagrange multiplier related to the second constraint
(Eq.~\eqref{eq::constraint2}). Since the number of votes is limited, $p(v|m)$
decays exponentially for $v\in [v_0,m/\Delta m]$ and it is zero otherwise. 

In order to verify if the distribution predicted by
Eq.~\eqref{eq::distribution} is compatible with real data, we consider the 2014
election for federal deputies in Brazil, using the dataset available in
Ref.~\cite{dataset2017}. Each state has its own ballot, with different
candidates and voters. Countrywide, this is an election with $6323$ candidates,
roughly $140$ million voters, and more than $280$ million dollars invested in
campaigns. We first analyze the results for the top four populated Brazilian
states, namely, S\~ao Paulo, Rio de Janeiro, Minas Gerais, and Bahia. These
states have each more than $10$ million voters and between $239$ (Bahia) and
$1364$ (S\~ao Paulo) candidates.  For each state, we grouped the candidates by
the amount of money that they reported to have spent in their campaigns.
Figure~\ref{fig::data1}A shows the average number of votes $\langle v\rangle$
received by a candidate as a function of the standard deviation $\sigma_v$ for
each money group. For most data point, the results are consistent with a linear
behavior (dashed line) as expected for an exponential distribution, where the
average and standard deviation are always equal. To verify the functional
dependence of the distribution, in Fig.~\ref{fig::data1}B shows the
distribution of votes, rescaled as $\bar{v}=(v-\langle v\rangle)/\sigma_v$,
where $\langle v\rangle$ and $\sigma_v$ is the average and standard deviation
of the number of votes per candidate in the same interval (logarithmic binning)
of money spent.  The distribution clearly follows the predicted exponential
behavior of Eq.~\eqref{eq::distribution} for more than $99\%$ of the
candidates.  However, for $\bar{v}>6$ the distribution deviates from the
predicted one (highlighted region in Fig.~\ref{fig::data1}B). There are nine
candidates in this region in the entire country, with five of them running in
S\~ao Paulo.  This is remarkable, as the theory predicts five in the entire
country and only one in S\~ao Paulo. This observation raises doubts about these
outliers and it could therefore call for a detailed analysis and validation of
their reported data about the campaign founding.

From the partition function~(\ref{eq::normal}), the average number of votes
received by a candidate that spent $m$ money in the campaign is,
\begin{equation}\label{eq::average_vote}
\langle v\rangle=v_0  + \frac{1}{e^{\mu}-1} + \frac{1-v_0+m/\Delta
	m}{1-e^{\mu(1-v_0+m/\Delta m)}} \ \ .
\end{equation}
The value of $\mu$ is obtained by imposing the second constraint
(Eq.~\eqref{eq::constraint2}) and considering $\Delta m$ as a free parameter.
Figure~\ref{fig::data2}A shows the number of votes per candidate against the
money spent in the campaign (gray circles) and the average value for candidates
in the same money group (orange circles), where the circles in blue correspond
to the outliers.  The solid line in Fig.~\ref{fig::data2}A is the non-linear
fit of Eq.~\ref{eq::average_vote} to the numbers of votes of all candidates as
a function of their financial resources, which gives $\Delta m =2.07$. As
shown, the excellent agreement between this fit and the averaged data points
extends over four orders of magnitude, with deviations found only for
candidates with very scarce resources, a fact that can be explained as follows.
For simplicity, we have considered that the minimum number of votes $v_0$ is
the same for all candidates, obtained by assuming that $v_0$ equals the average
number of votes for candidates who spent less than $1200$
dollars~\cite{Melo18}. In general, however, every candidate has a different
$v_0$, depending on several factors such as, his/her party, visibility, and
social status. 

Through Eq.~\eqref{eq::distribution}, it is also possible to predict, from the
reported amount of money spent by each campaign, the distribution of votes for
an election. As an example, let us consider again the election in the state of
S\~ao Paulo. Figure~\ref{fig::data2}B shows the distribution of the number of
votes (red dots) obtained by each candidate. To predict the distribution of
votes for this election, we assigned randomly a number of votes to each
candidate from a distribution given by Eq.~\eqref{eq::distribution}, with $m$
equal to the amount of money spent in the campaign, as declared by the
candidate. The solid line in Fig.~\ref{fig::data2}B is the predicted outcome,
which is in excellent agreement with the empirical data. Once more, in the
proposed framework, $\Delta m$ is a free parameter that relates to the amount
of money spent by a candidate campaign and the maximum number of votes that it
can receive. Its value was estimated by fitting Eq.~\eqref{eq::average_vote},
as explained before.

\textit{Conclusions.} We have shown, using the principle of maximum entropy,
that the distribution of votes received by a candidate should follow an
exponential distribution parameterized by the amount of money that was spent
in her/his campaign. This prediction is consistent with real data from a very
large proportional election, with $6323$ candidates. Furthermore, as the money
spent in a campaign is heterogeneously distributed among candidates, we
developed a framework based on superstatistics to establish the relation
between the distribution of money spent and of votes. Within this framework, it
was possible to predict the outcome of a ballot from the distribution of money
spent, and identify potential cases of misconduct either in the report of
fundraising and spending or on vote counting.

For several proportional elections, the distribution of votes per candidate is
fat tailed~\cite{Chatterjee2013}, what has motivated an enthusiastic discussion
about the underlying mechanism~\cite{Castellano2009}. As our theoretical
approach shows, for an election, if all candidates spent the same amount of
money in their campaigns, the expected distribution of votes would be
exponential. So, the fat-tailed distribution is a consequence of an
heterogeneous distribution of resources.  This is consistent with the reported
power-law distribution of money spent by candidates in the same
elections~\cite{Melo18}.

\begin{acknowledgments}
 We thank the Brazilian Agencies CNPq, CAPES, FUNCAP and FINEP, the
	FUNCAP/CNPq Prunes grant, and the National Institute of Science and
	Technology for Complex Systems in Brazil for financial support.  NA
	acknowledges financial support from the Portuguese Foundation for
	Science and Technology (FCT) under Contract no. UID/FIS/00618/2013.
\end{acknowledgments}

\bibliography{paper}

\end{document}